# Novel synthesis of topological insulator based nanostructures ($Bi_2Te_3$) demonstrating high performance photodetection

*Alka Sharma[1,2], T D Senguttuvan[1,2], V N Ojha[1,2] and Sudhir Husale[1,2]**

[1]*Academy of Scientific and Innovative Research (AcSIR), National Physical Laboratory, Council of Scientific and Industrial Research, Dr. K. S Krishnan Road, New Delhi-110012, India.*

[2] *National Physical Laboratory, Council of Scientific and Industrial Research, Dr. K. S Krishnan Road, New Delhi-110012, India.*

**E-mail:* husalesc@nplindia.org

**Abstract**

The rapid progress in 2D material research has triggered the growth of various quantum nanostructures- nanosheets, nanowires, nanoribbons, nanocrystals and the exotic nature originating through 2D heterostructures has extended the synthesis of hybrid materials beyond the conventional approaches. Here we introduce simple, one step confined thin melting approach to form nanostructures of TI (topological insulator) materials, their hybrid heterostructures with other novel 2D materials and their scalable growth. The substrate and temperature dependent growth is investigated on insulating, superconducting, metallic, semiconducting and ferromagnetic materials. The temperature dependent synthesis enables the growth of single, few quintuples to nanosheets and nanocrystals. The density of nanostructure growth is seen more on fabricated patterns or textured substrates. The fabricated nanostructure based devices show the broadband photodetection from ultraviolet to near infrared and exhibit high photoresponsivity. Ultimately, this unique synthesis process will give easy access to fabricate devices on user friendly substrates, study nanostructures and scalable growth will enable their future technology applications.

## Introduction

At nanoscale region, nanostructures compared to their bulk counterpart show quantum confinement effects resulting in exotic physical, and electronic transport properties. The growth of quantum nanostructure is highly desirable if the method is cost effective, capable of achieving scalable synthesis, growth on device friendly substrates and also applicable to form the heterostructures of novel materials. Since last few years, many 2D layered materials such as graphene, $MoS_2$, $WS_2$, BN, $Bi_2Se_3$, $Bi_2Te_3$ etc. have demonstrated exotic properties and showed their potential nanodevice applications. Among these the topological insulator, a new exotic state of matter, shows robust metallic surface states protected by time reversal symmetry which has attracted an immense interest in synthesis of these quantum nanostructures due to their potential applications in spintronics[1,2], thermoelectrics[3], dissipationless electronics[4], hosting Majorana fermions for quantum computation studies[5], broadband photovoltaic[6], infrared detectors[7], photodetectors[8-10] etc. The transport through topological surface states are often hindered due to the contribution coming from bulk of the material but quantum materials in 2D or 1D effectively enhance the surface transport suppressing the bulk contributions. The growth of these nanostructures can be achieved by using techniques such as molecular beam epitaxy (MBE)[11,12], vapor liquid solid (VLS)[13,14] or CVD[6], vapor solid (VS)[15,16], metal-organic chemical vapor deposition (MOCVD)[17,18], pulsed laser deposition (PLD)[19,20] and solvothermal methods[21,22] but most of such techniques are either expensive, need lattice matching, precursor molecules, chemicals, face great difficulty to form heterostructures and more importantly substrate used for deposition are not device friendly. Exfoliation[23,24] (mechanical or hydrothermal) is another simple approach but yield is often less and depositions of definite shapes of nano structures such as single quintuples, hexagons, nanowires are almost impossible.

Moreover the heterostructures of these novel materials when combined with other 2D materials give birth to many interesting puzzles and phenomena. Hence there is a great demand to synthesize a material from two or more materials aiming exciting desired properties and functionalities. But synthesis of such hybrid nanostructures is a complicated process for the present synthesis techniques.  There is no general and simple solution for the fabrication of heterostructures and one step solution is needed for many applications that include photodetectors, biosensors, thermoelectric devices, solar cell, field effect transistors etc. Here we report simple catalyst, precursor, liquid free method which can synthesize these quantum nanostructures, their hybrid forms with high yield and offer a scalable approach on a device friendly substrate. Scalable manufacturing of van der Waals heterostructures might be possible with this method similar to vapor solid method which allows direct growth of topological insulator based nanostructures on 2D layered materials. This is a unique method compared to other techniques (supplementary information Table 1) and can be employed for synthesizing of other topological insulator based nanostructures and heterostructures with novel 2D materials.

## Results and Discussion

Figure 1 shows the experimental schematics of confined thin film melting and substrate dependent growth of various nanostructures.  Sections (I and II) represent the cross sectional view and simple steps of this method. Section (II A) represents the morphologically different types of substrates (s1-s4) used in this study and B is the sputter coated thin film.  The confined melting process shows the top (growth) substrate (A) is in contact with the down (source) substrate (B) where sputtered deposited thin film of topological insulator ($Bi_2Te_3$) was mounted on a temperature controlled heater. The assembly was kept in vacuum chamber at pressure $10^{-5}$ mbar or better. After heating for > 30 min the substrate assembly was cooled down to room temp. The molten species of Bi and Te were formed and simultaneously diffused on top substrate where these species get adsorbed and nucleate to form the nanosheets as shown in the figure. The substrates were detached at room temp and were used as it is for FESEM characterization. The growth of nanostructures, their shapes, sizes and density were found dependent on the heating temp and type of the substrate used for the growth (part-IV) which we have experimentally studied here.

**Growth on Device Friendly Substrates**

The FESEM images shown in Figure 2 (a-h) represent the growth of $Bi_2Te_3$ nanostructures on insulating substrates silicon oxides and nitrides. Similar growth on other insulating substrate such as glass, quartz, sapphire are shown in supplementary information Figure S1. Various thin nanowire or nanoribbon type of structures are visible in Figure 2 (a). The as grown nanosheets having shapes like hexagon, triangles are shown in Figure 2 (b & c). Growth of single (light contrast) and few quintuples (bright contrast) of $Bi_2Te_3$ are shown in Figure 2 (d). The thickness of the thinnest nanosheet was determined using AFM which is shown in the inset (Figure d) and the step height of ~ 1nm indicates the single quintuple layer. Individual small dots, probably the nucleation points, were routinely observed at many places at the center of the single or few quintuple layers (supplementary information Figure S2). Such nucleation center was also observed previously during the deposition of $Sb_2Te_3$ quintuples by vapor solid method.[25] The energy dispersive X ray (EDS) analysis is shown in the insets of the Figure 2(e) which shows the presence of bismuth and tellurium elements with atomic compositions of about ~ 24.7 and 33.8 %, respectively. The elemental mapping confirms the uniform distribution of these elements. The nanosheets, nanowire or nanoribbon type of structures grown on silicon nitride substrates are clearly visible in Figure 2(f-h). The nanosheets were randomly distributed on these substrates. The lateral sizes of the nanosheets were observed from few hundreds of nanometer to microns and thickness was estimated using AFM (inset figure 2f) . Many places we observed hexagonal, triangular morphologies with top flat surface indicating high quality growth which is important for device fabrication and quantum phenomena based basic studies. The high crystalline quality of the material was observed and characterized by using HRTEM and Raman techniques which are shown in supplementary information Figure S3 and S4 respectively.

**Growth on 2D layered materials**

The coherent stacking of two different novel 2D materials forms the new exciting electronic system with unusual properties, phenomena and unique functionalities[26-28]. But the synthesis of TI nanostructures and their heterostructures with novel 2D materials is rare due to the complexity in the preparation of such quantum materials. Growth of $Bi_2Te_3$ nanostructures forming heterostructures with other 2D layered materials such as graphene, $MoS_2$ and $Bi_2Se_3$ is shown in Figure 3. Micromechanically exfoliated flakes of 2D layered materials were deposited on $SiO_2$/Si substrate and same were used as to form the heterostructures. Figure 3(a) and its inset represent Frank –van der Merwe type growth of $Bi_2Te_3$ hexagonal nanosheets on graphene flakes. Growth shows random coverage with higher affinity towards the edges. At some places, these deposited nanosheets further act like seeds for the vertical growth of $Bi_2Te_3$ nanosheets (Stranski – Krastanov ) as shown in Figure 3(b). When thin film was melted at temp $\geq T_m$ (melting temp of $Bi_2Te_3$), we find 3D type growth of nanocrystals (Volmer –Weber growth) indicating the less nucleation density and rapid growth as shown in Figure 3(c). Here melting at high temp might provide the rapid interaction of adatoms with initial adsorbed molecule on the substrate. The flower type of growth was observed on $Bi_2Se_3$ nanosheets (Figure 3d) whereas shapes like triangles or hexagons were distinguishable on the $SiO_2$ substrate as shown by the arrows in Figure d. Note that bismuth selenide is also a low temp melting material and any distortion formed in lattice during heating may affect the nucleation and diffusive growth process. And here, we have observed the growth of large sheets without sharp edges covering most of the substrate flake area. The growth of $Bi_2Te_3$ nanostructures on $MoS_2$ is shown in Figure 3(e-h). Similar to graphene substrate, growth of hexagons (Figure 3e) at temp $< T_m$ and nanocrystals (Figure 3f) at temp $\geq T_m$ were observed.

The edges showed high coverage growth and vertical growth was also found reproducible. The growth of very tiny nanostructures was also noticed on few flakes (Figure 3g&h).

**Scalable growth on ITO substrate**

Above results indicate that the growth of nanostructures is not uniform on flat insulating and 2D layered material substrates. Growth of quantum nanostructures on large area with high density and high reproducibility could be the key for many quantum technologies in future. The ITO coated glass shows very nice textured film with nano sized grain domains and boundaries as visible in Figure 3(i) and inset of I shows AFM topographic image for the same. The growth area is clearly visible as compared to bare ITO substrate in Figure 3(i). Insets of I and II represent AFM images of the bare ITO substrate and $Bi_2Te_3$ nanosheets grown on these substrates, respectively. Roughness of the textured film was around ~ 5 nm and the thickness of the nanosheets was estimated about ~15 nm from the AFM images. Fig 3(j and k) show the substrate triggered, high density growth of various $Bi_2Te_3$ nanosheets. Figure 3(l-n) show similar growth and densely packed nanosheets are clearly visible along with some vertically grown structures i.e. growth of the nanostructure perpendicular to the substrate. The lateral dimensions are less than micron size indicating the dangling bonds present on this substrate could restrict the lateral growth of the nanosheet and adatoms might prevent the lateral growth of the nanosheets and prefer vertical growth at some places. This indicates that the neighboring quintuples do not show van der Waals bonding. Important to note that the growth was observed over a large area, shown in the inset of Figure 3(l), demonstrating scalability of this method which may find potential use in technological applications because of cost effectiveness and dense coverage over a large area.

**Growth on lithographically patterned substrates**

All the above results clearly show that substrate morphology play an important role in the growth of $Bi_2Te_3$ nanostructures. To study the density of the nanostructure growth and their affinity compared to flat substrate, we purposefully fabricated different types of patterns on insulating substrates and found interesting growth of $Bi_2Te_3$ nanostructures on these fabricated patterns. The growth of $Bi_2Te_3$ nanostructures on ebeam fabricated NiFe nanostrips is shown in Figure 4 (a-c). Since the edges give rise to high surface energy, more specific growth is expected on such patterns. Compared to substrate, high growth affinity on the nanostrips was clearly observed in Fig 4(b –c). Insets (I &II) in Figure 4(c) show elemental mapping of Bi and Te, respectively, indicating the uniform distribution of the material. Topological quantum nanostructures integrated with superconducting material possess exciting exotic properties. Dots of Niobium were fabricated using ebeam lithography (Figure 4d) and we found high density growth of $Bi_2Te_3$ nanostructures on these dots (Figure 4e). High density growth was also observed on lithographically fabricated squares of Titanium (Ti) material (Figure 4f) and insets I and II represent the elemental mapping of Bi and Te, respectively. More location specific systematic growth and arrangement of $Bi_2Te_3$ nanostructures were observed on fabricated patterns such as nested loops, arrays of squares, dots arranged in circular manner and they are shown in Fig 4(g-i), respectively. Here similar to vapor solid growth, the presence of an impurity- dust, scratches, grains and grain boundaries, implanted ions etc. serves as a precursor and strongly helps during the growth of nanowires or nanosheets. We have observed that ion implantation i.e. pattern milled by FIB using Ga ion enhances the deposition of nanostructures (Figure 4j-l). Note that the ion implantation also induces the additional compressive stress at localized area on the substrate and favors the atomic diffusion and growth of nanowires. Ion stress induced growth of nanowires on patterned area was studied earlier[29]. Single or less dense nanowires can be easily found with this method which is good for the device development research and the growth on patterned substrate is attractive and could be used as an alternative approach.

**Temperature dependence on the growth of nanostructure**

The Figure 5 (I) represents the temperature dependent evolution of different $Bi_2Te_3$ nanostructures observed in our study. The noticeable growth of nanowires, nanoribbons were routinely observed at temp ~ 350° C ( ± 50) along with very few thin nanosheets and the density of nanostructure growth was always dependent on the nucleation sites, defective and engineered places. Note that compared to Si/$SiO_2$/ITO substrates, bismuth telluride has a higher thermal expansion coefficient (TEC). Thus the sample heating impedes the expansion due to mismatch in TECs and induces the microscopic compressive stresses in $Bi_2Te_3$ film. The growth of $Bi_2Te_3$ nanowires due to the stress release or induced mass flow has been observed in sputtered deposited polycrystalline films[30] and thermal treatment used for the growth of CuO nanowires[31]. During heating, due to compressive stress, the mobile atoms e.g. Bi and Te preferably diffuse and accumulate at grain boundaries. Under cooling conditions, film undergoes tensile stress and diffused atoms move upward direction at the interface of $Bi_2Te_3$ thin film and may find an energetic nucleation sites for the growth of nanowires at the top substrate.

For the temp range ~500° C ( ± 50), we mostly observed nanosheets of various shapes and sizes and very few nanowires or nanoribbons. The very high density of growth was noticed on textured (ITO coated glass) substrates at similar temperature. At this temperature, semi-molten phases could occur at many places in $Bi_2Te_3$ film which can be called as hot spots. During the cooling process probably a transformation in crystallization happens and tensile stress push might help for molecule adhesion, nucleation and subsequent growth at the top substrate. In case of ITO substrate, during cooling, molecules find many adhesion sites present on the textured ITO film where high density of growth was observed. During cooling/solidification, granular structure and grain boundaries act as substrates for the nucleation of the nanosheets by lowering the activation barrier and the closely spaced grains helps the simultaneous nucleations favored by low surface energy over a large surface. The growth and its dependence on the grains and grain boundaries present in the thin films are not investigated here. Ramping the temperature beyond ~ 650° C ( ± 50), we observed very consistent growth of thick nanocrystal like structures along with some few layered nanosheets on substrates like $SiO_2$, $Si_3N_4$, Graphene, $MoS_2$ or ITO.

**Optoelectronics measurements**

Further we have investigated the broadspectral photodetection properties of a $Bi_2Te_3$ nanowire grown by this method (Figure 5 (II)). Two probe optoelectrical measurements were carried out using Cascade Microtech instrument accompanied with Keithley 2634B source meter and light sources - UV-325 nm ($P_d$ = 1 to 13 mW/cm$^2$), Visible - 532 nm ($P_d$ = 32 mW/cm$^2$), NIR-1064 nm ($P_d$ = 1 to 18 mW/cm$^2$) and 1550 nm ($P_d$ = 1 to 10 mW/cm$^2$). The nanowire was biased at constant 1V and sudden increase in the current was observed due to the light illuminations of different wavelengths. We have observed the photocurrent dependency on the incident laser light illumination and the systematic increase in photocurrent with increase in laser power density was studied using the relation $I_{ph} = P^{\theta}$. The photoresponsivity of the $Bi_2Te_3$ nanowire was estimated using relation, $R = \frac{I_{ph}}{P_d \cdot A}$, where, R is the photoresponsivity, $I_{ph}$ is the photocurrent, $P_d$ is the power density and A is the active area of the $Bi_2Te_3$ nanowire device. The maximum photoresponsivity of about ~ 286 A/W is observed for the nanowire at 1550 nm, which shows high performance compared to the earlier reported photoresponsivity of the $Bi_2Te_3$ based photodetectors[32-36]. Further the estimated rise, decay time constants and detectivity were observed about 250 ms, 195 ms and 6.6.x $10^9$ Jones respectively.

## Conclusion

In conclusion, we report synthesis of topological quantum materials and their hybrid forms on planar substrates like SiN/SiO2/ ITO by a very simple cost effective technique eliminating the need of expensive instruments, precursors or catalyst. Compared to expensive or complicated methods, one step synthesis (assuming the availability of source and growth substrate) is simple and more attractive because the results show that the growth of bismuth telluride nanostructures (quintuples, nanosheets, nanoribbons and formation of heterostructures) can be achieved very easily on a device friendly substrate which makes this method more appealing for research in device fabrication and transport studies. The density of nanostructures and various shapes can be tuned using the substrate and the growth temperature respectively. The data clearly shows the presence of Volmer–Weber (nanocrystals or thicker sheets), Frank–van der Merwe (2D nanosheet, layer by layer), and Stranski–Krastanov (single quintuple plus islands, mixed– layer and island) growths. This method can be universalized for other low melting temp based topological insulator materials and can form hybrid heterostructures with ease compared to other reported methods. High density nanosheet growth on textured film over a large area shows the scalability and the assembly of quantum nanostructures ready for future technological applications. Our experimental approach is simple but both the top and down substrates are in contact with each other, hence, for a clear understanding behind our results we need more theoretical and probably experimental designs in future. However the exotic nature of topological insulator based quantum nanostrcutures and their 2D hybrids may give new directions to synthesize the nanostructures and could find promising applications in nanoelectronics, optoelectronics, sensing, catalysis, energy storage and thermoelectrics.

# Methods

**Synthesis and electrical characterization of $Bi_2Te_3$ nanostructures.** The substrates used in this study (silicon nitride, silicon oxide, quartz, sapphire, glass) were ultrasonicated in acetone and subsequently cleaned with acetone, isopropanol, methanol and DI water. Oxygen plasma treatment was performed prior to deposition of thin films and growth of nanostructures on these substrates. Source substrate (thin film of $Bi_2Te_3$) was made using sputtering technique and thickness of the film was optimized using AFM. After deposition of $Bi_2Te_3$ thin film, growth substrate was placed directly top on the sputtered thin film and was hold together using a gentle clamp. Heating was started once the vacuum of the chamber was in the range of about ~ 5 x $10^{-7}$ mbar or better. After heating substrates were cooled down to room temp and growth substrate was characterized using FESEM and optical microscopy. For device fabrication, big metal contact electrodes of Au ~ 80 nm and Ti ~ 10 nm were deposited on the growth substrate using shadow mask. Pt electrodes were directly deposited on the $Bi_2Te_3$ nanostructures using gas injection system (FIB attachment) and connected to big metal pads with the same technique. Electrical measurements were carried out in the probe station using source meter 2634B and laser light sources.

**Data availability.** All experimental data required to evaluate and interpret the conclusions are present in the main manuscript or supplementary materials file. Additional data or information related to this paper may be requested from the corresponding author (*E-mail: husalesc@nplindia.org).

### Acknowledgments

We thank CSIR NPL for financial support, Mr. Sandeep Singh for the AFM imaging, Mr. Dinesh for HRTEM characterization, Dr. Govind for UV laser support and Dr. Sangeeta Sahoo for critical reading and comments.

### Author Contributions

A.S. performed the synthesis of nanostructures using sputtering system, carried out all the optoelectronic measurements, analysed HRTEM and Raman characterization data, schematics, figures and fabricated the nanodevices with SH. T.D.S. and V.N.O provided FIB tools, operational support and materials. S.H. conceived, designed the overall experimental strategy, supervised the research and wrote the manuscript. All the authors read and commented on the manuscript.

## Additional Information

**Supplementary information** accompanies this paper at.

**Competing interests:** The authors declare that there are no competing financial or non-financial conflicts of interests.

**Figure 1**

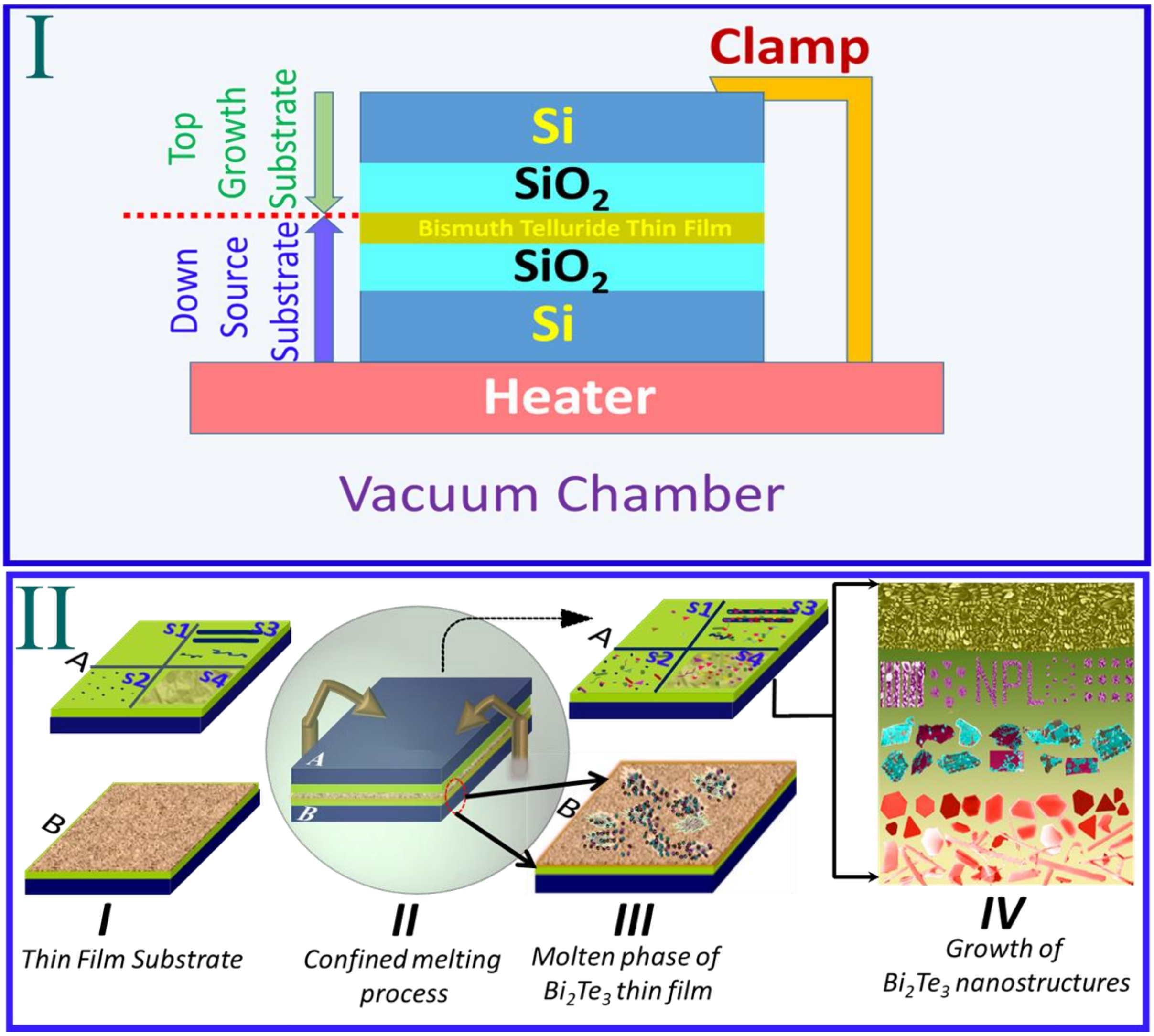

**Figure 1.** Schematic of the confined thin film melting method (cross sectional view I) and different steps (II). (II A) indicates the different types of substrate used for the growth of $Bi_2Te_3$ nanostructures**.** (B) shows the thin film deposited substrate.

**Figure 2**

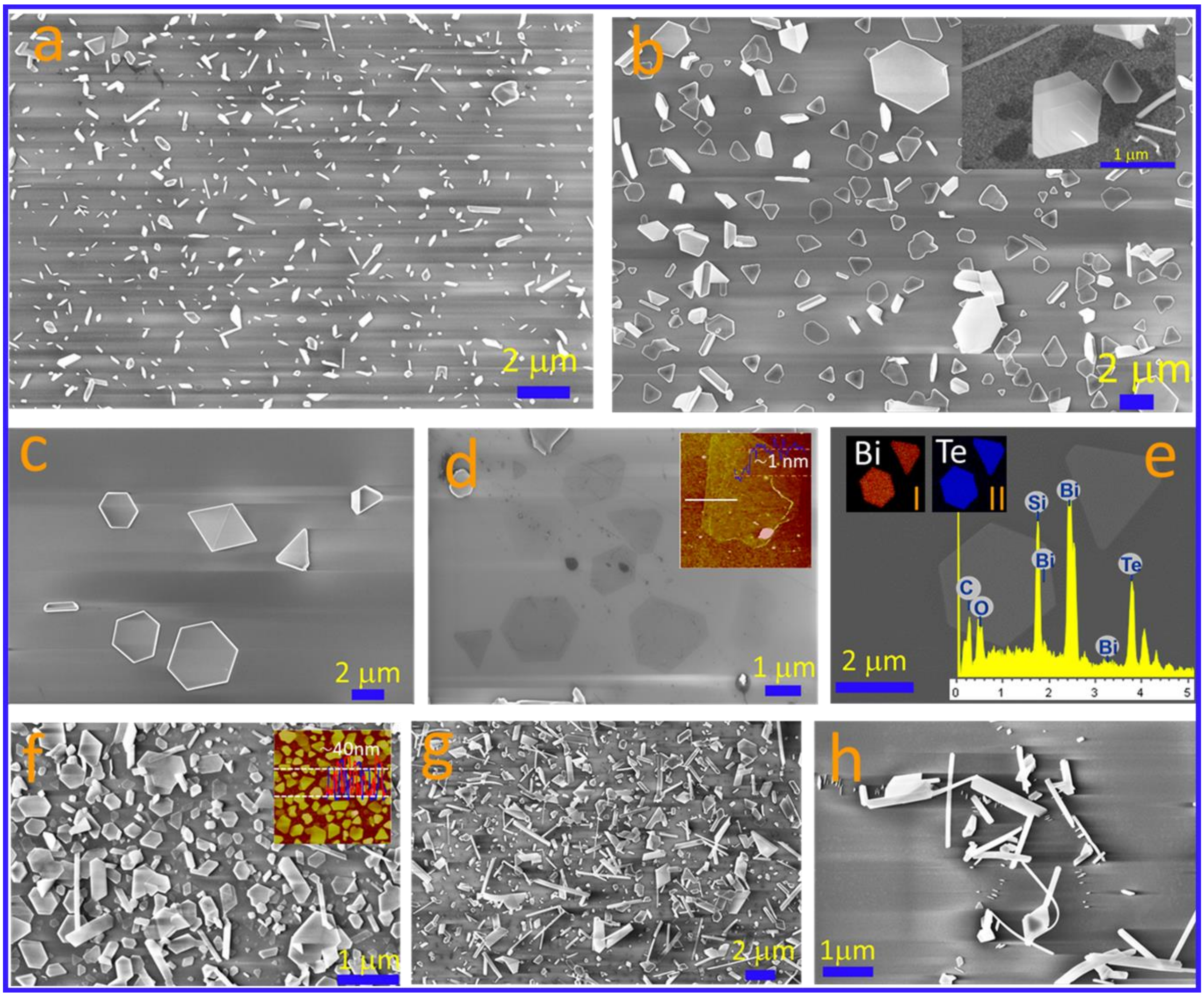

**Figure 2 Growth on device friendly substrates ($SiO_2$ and $Si_3N_4$).** Fig (a-h) FESEM images representing a) short nanowire and nanosheet like structures, b&c) nanosheet dominated growth, inset (b) indicates the growth of different layers. Very thin nanosheets are visible in Fig (d), AFM image (inset) estimates the growth of single quintuple. Fig (e) shows the EDS analysis of nanosheets. Fig (f-h) nanostructure growth on silicon nitride substrate, inset (f) shows the height measurement using AFM image.

**Figure 3**

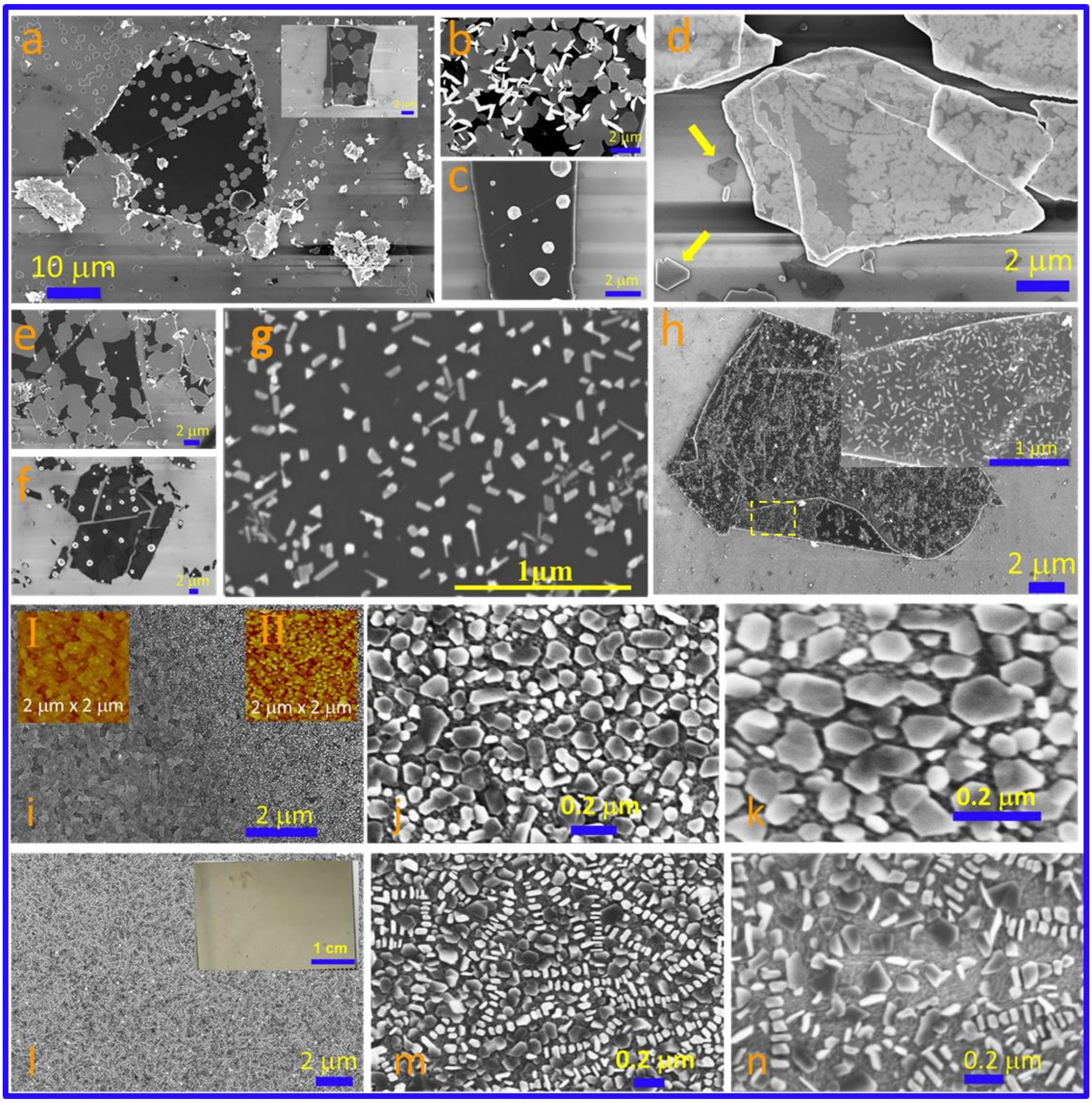

**Figure 3**. Heterostructure and scalable growth. (a-h) FESEM images demonstrating growth of $Bi_2Te_3$ nanostructures on graphene (a-c), $Bi_2Se_3$ (d) and $MoS_2$ flakes (e-h). Figure (i-n) large area growth of $Bi_2Te_3$ nanostructures on ITO coated glass substrate. Inset I&II (Fig i) represent the AFM image of ITO substrate and $Bi_2Te_3$ nanostructures on ITO substrate respectively. Inset (Figure l) represents the optical image demonstrating the possibility of scalable growth.

**Figure 4**

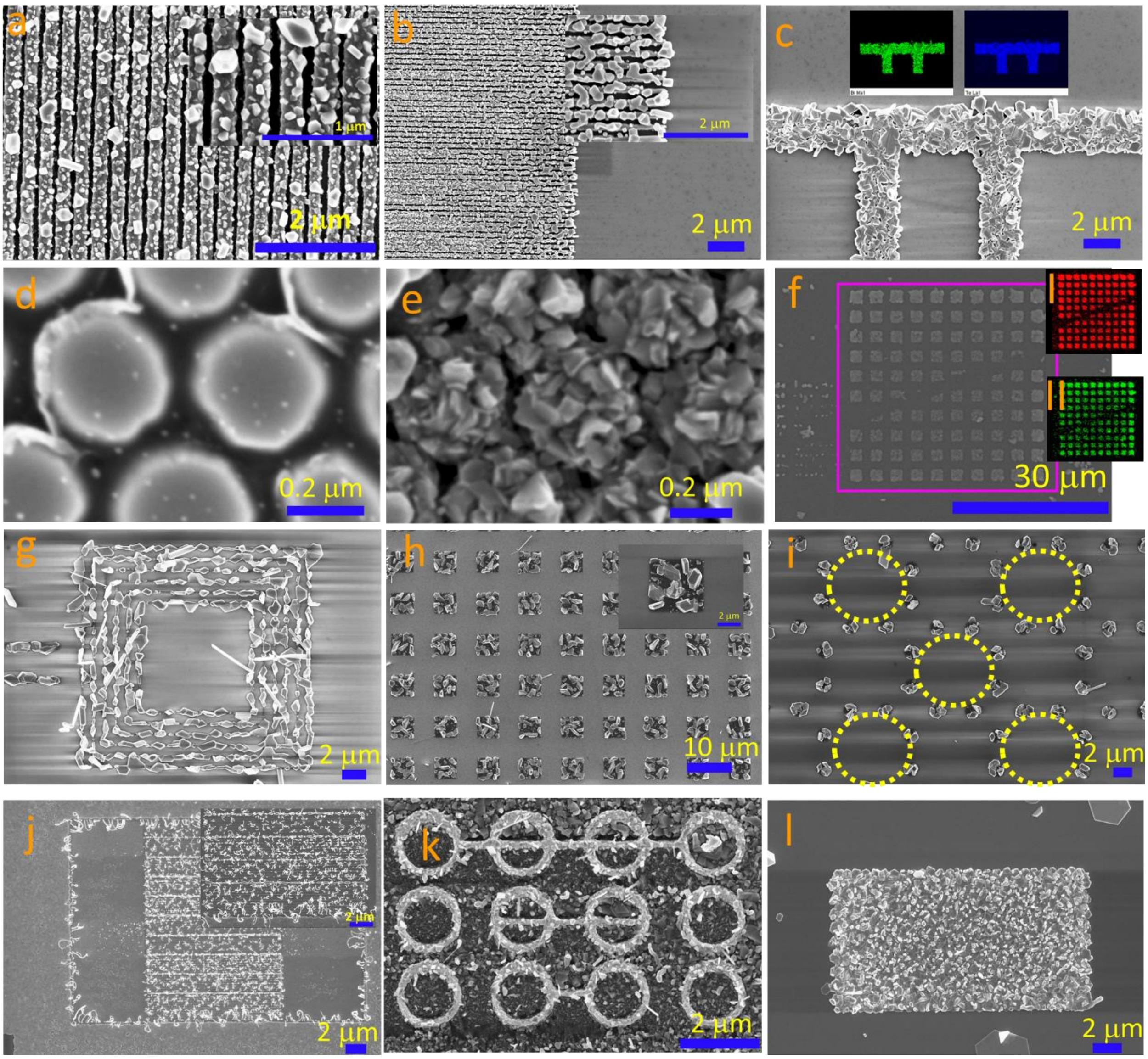

**Figure 4.** Growth of $Bi_2Te_3$ nanostructures on ebeam fabricated patterns (a-i) and FIB milled patterns (j-l). Inset I&II in Figure c and f show the elemental mapping obtained using EDS.

**Figure 5**

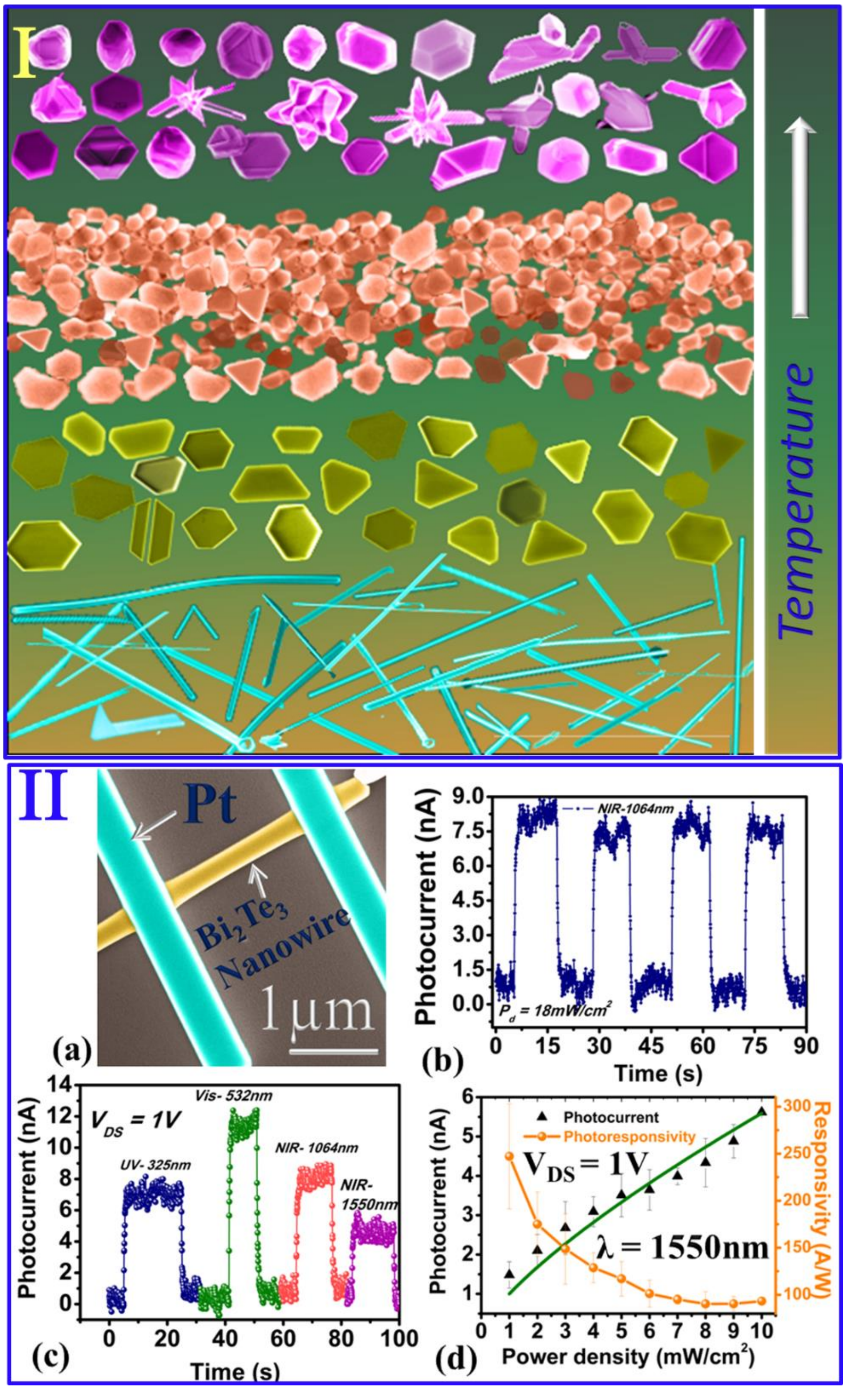

**Figure 5.** Temperature dependent growth of various $Bi_2Te_3$ nanostructures observed in this study and optoelectronic characterization of a $Bi_2Te_3$ nanowire device. (a) The false colour FESEM image of nanodevice. (b) The stability of nanowire device is examined under 1064 nm illumination. (c) Spectral dependent photocurrent–time domain of the nanowire at bias voltage 1V. (d) The power density dependent photocurrent and responsivity of the $Bi_2Te_3$ nanowire device. The green curve shows the power law fitting of photocurrent at different power densities from 1 to 10 mW/cm$^2$.

Supplementary Information

# Novel synthesis of topological insulator based nanostructures ($Bi_2Te_3$) demonstrating high performance photodetection

Alka Sharma[1,2], T D Senguttuvan[1,2], V N Ojha[1,2] and Sudhir Husale[1,2]*

[1]Academy of Scientific and Innovative Research (AcSIR), National Physical Laboratory, Council of Scientific and Industrial Research, Dr. K. S Krishnan Road, New Delhi-110012, India.

[2] National Physical Laboratory, Council of Scientific and Industrial Research, Dr. K. S Krishnan Road, New Delhi-110012, India.

*E-mail: husalesc@nplindia.org

Supplementary Information includes:

1. Fig S1, $Bi_2Te_3$ nanostructure growth on glass, quartz and saphhire

2. Fig S2. Growth of $Bi_2Te_3$ quintuples, dots and nanocrystals

3. Fig S3 HRTEM characterization of $Bi_2Te_3$ nanosheets

4. Fig S4 Raman characterization of $Bi_2Te_3$ nanosheets

5. Fig S5 Deposition efficiency on $SiO_2$, Au, $SI_3N_4$, ITO, ebeam patterned and FIB milled substrates

6. Fig S6 Broadspectral photoresponse of different nanostructure devices.

7. Fig S7. Rise and decay time curve fitting

8. Table I. Synthesis comparison of topological insulator nanostructures using various techniques

9. Table II. Photorepsonse properties of topological insulator based nanostructures and thin films.

1. Fig S1. $Bi_2Te_3$ nanostructure growth on glass, quartz and saphhire

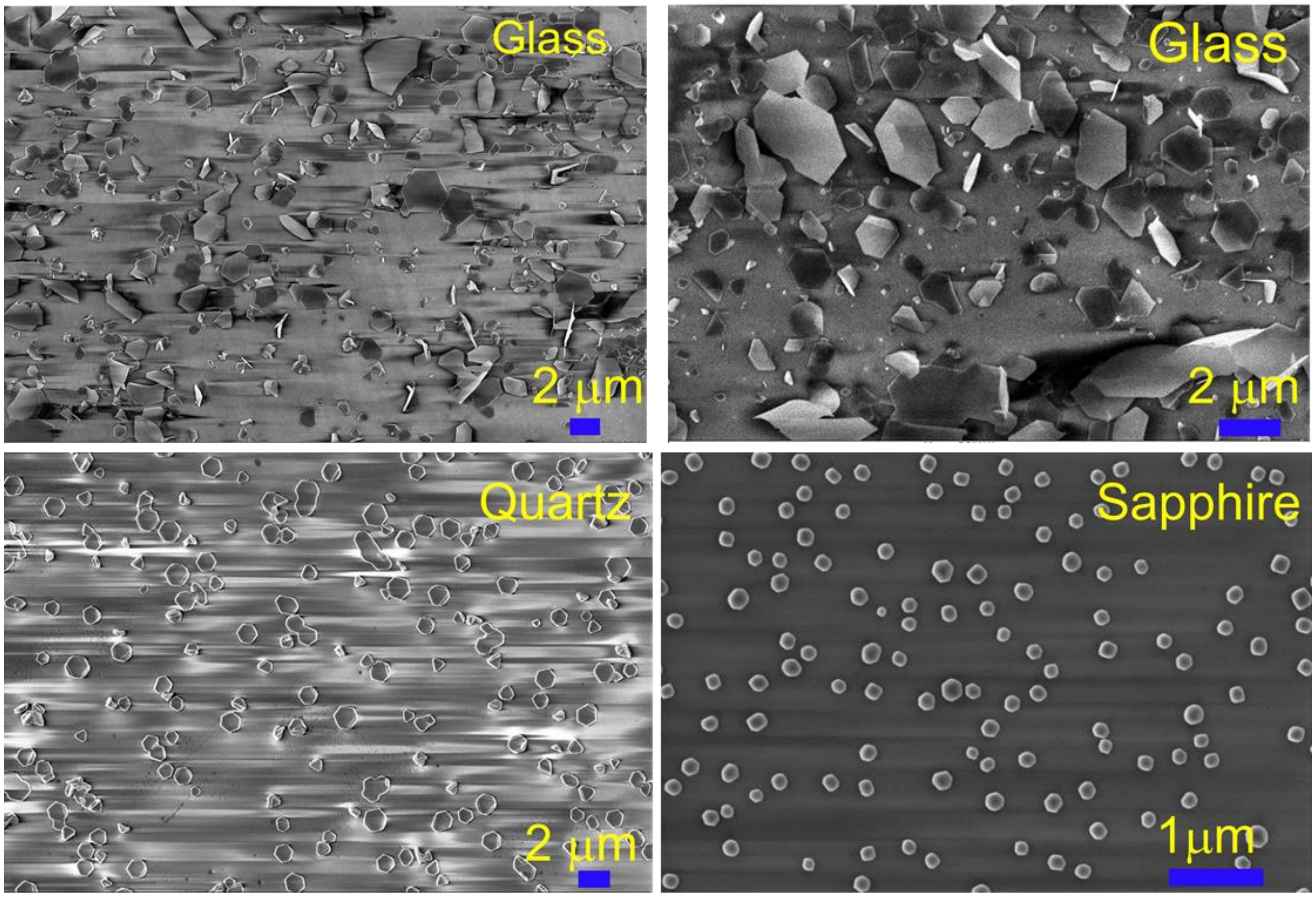

2. Fig S2. Growth of $Bi_2Te_3$ quintuples, dots and nanocrystals

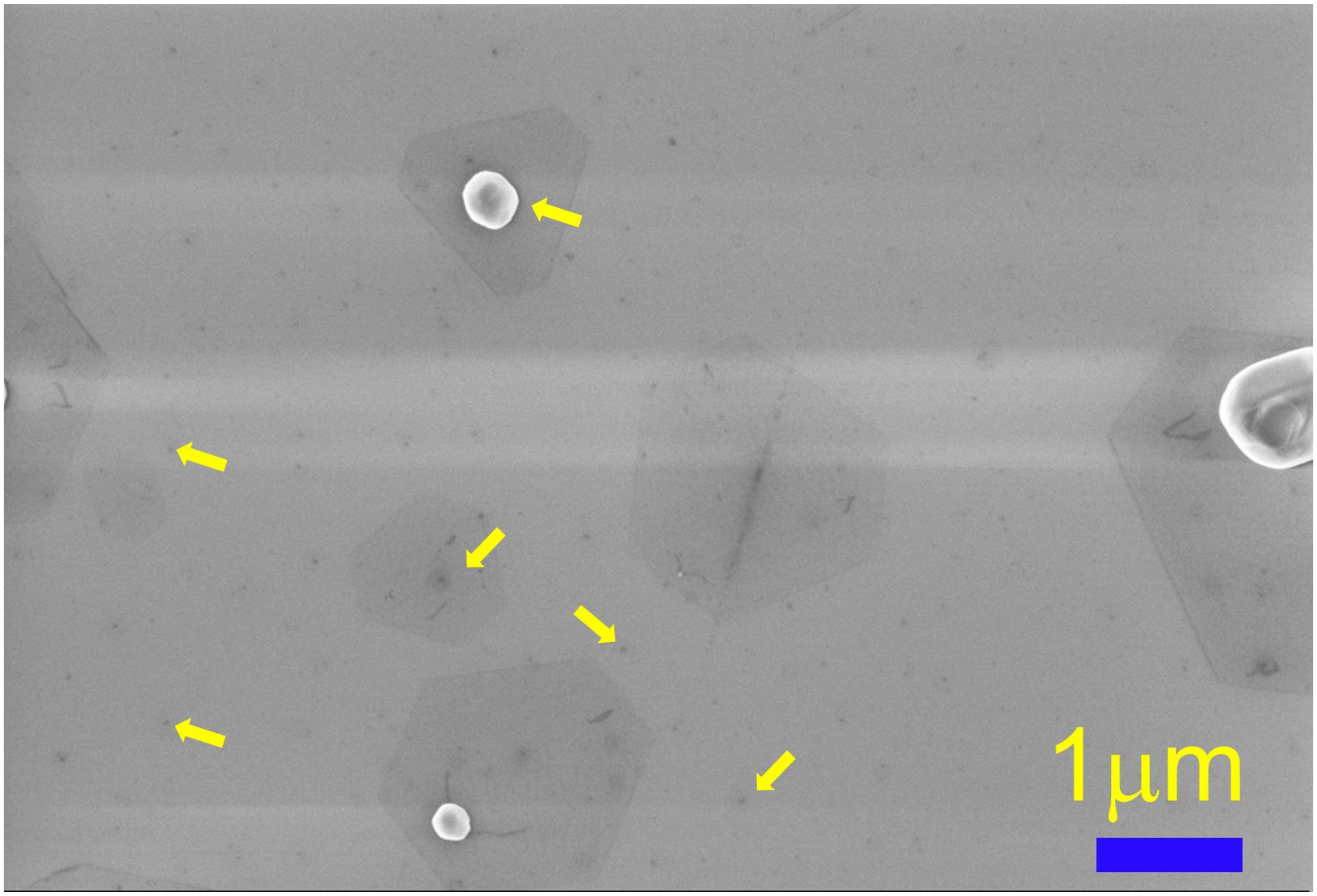

3. Fig S3. HRTEM characterization of $Bi_2Te_3$ nanosheets

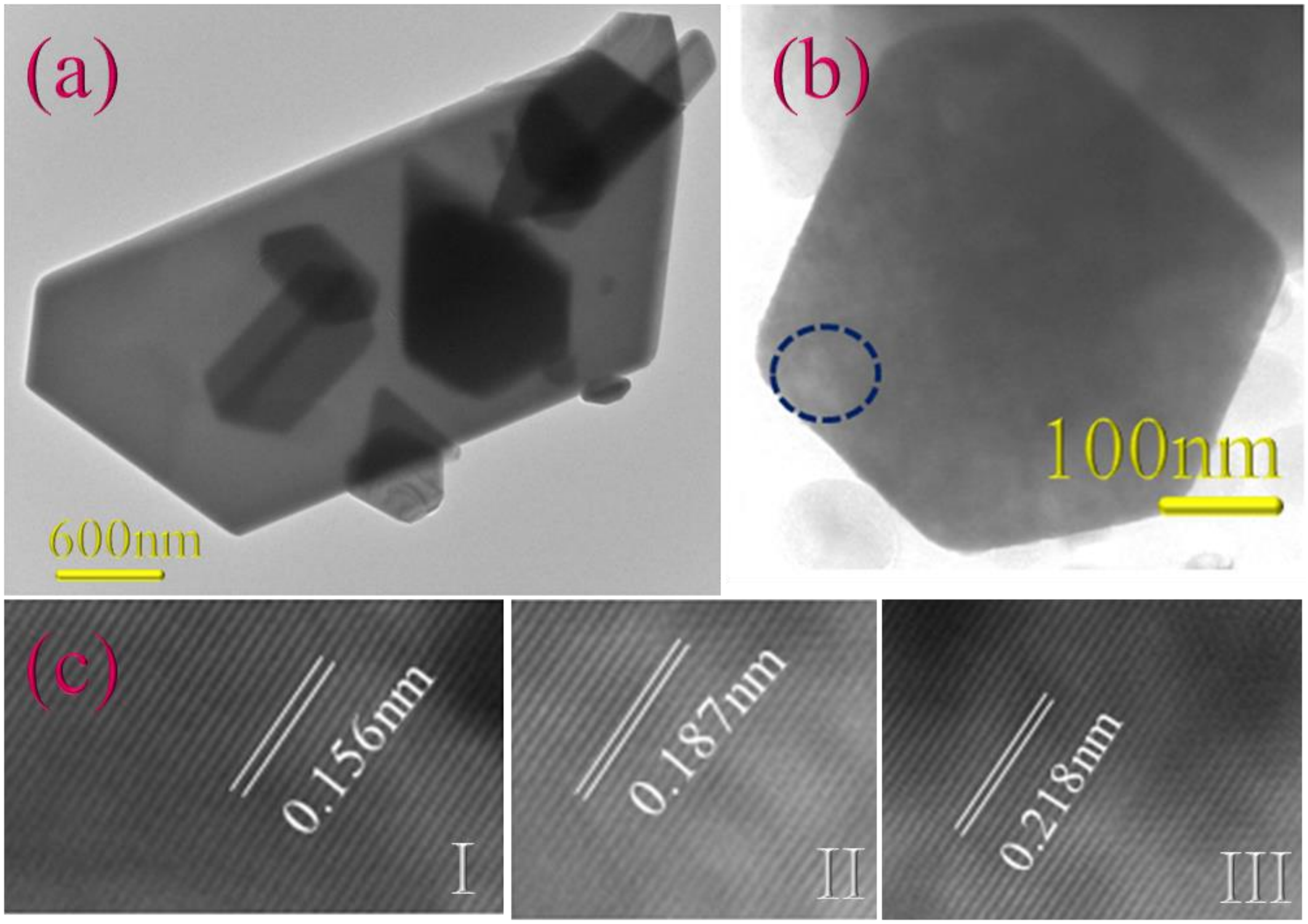

*Fig.S3 HRTEM images of $Bi_2Te_3$ nanosheets. (a,b) Bright field high resolution micrograph of the $Bi_2Te_3$ nanostructures showing different shapes- hexagon, triangle, pentagon etc. (c I-III), atomic scale images of the location shown in Fig.S2 (b) by dotted circle.*

The high resolution transmission electron microscopy (HRTEM, model Tecnai G2F30 STWIN) was used to determine the crystalline nature of $Bi_2Te_3$ nanostructures. HRTEM images shown in S3 (a) indicate that the nanostructures possess sharp edges and elongated or flat morphologies. The HRTEM micrograph of $Bi_2Te_3$ nanostructure clearly shows the atomic planes of $Bi_2Te_3$ with interplaner distance (d) 0.156, 0.187 and 0.218 nm.

4. Fig S4. Raman characterization of Bi2Te3 nanosheet

Following fig shows the Raman spectrum of the $Bi_2Te_3$ nanosheet grown on Ti substrate. The three characteristics vibration modes peaks of $Bi_2Te_3$ nanosheet are labelled with $A_g^1$ , $E_g^2$ and $A_g^2$.

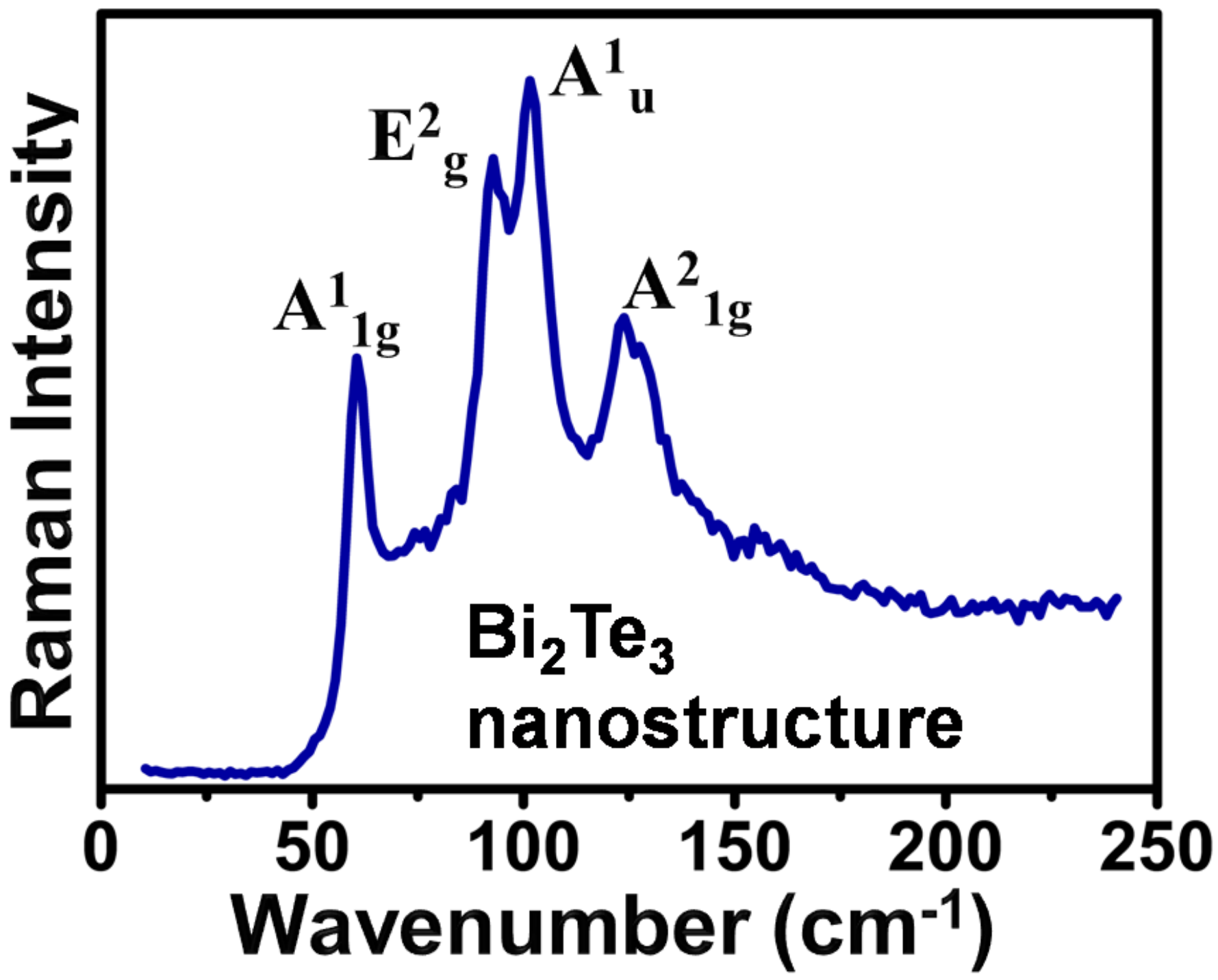

*Fig.S4 Raman spectrum of hexagon shaped $Bi_2Te_3$ nanostructure grown on Ti substrate with vibration modes $A^1{}_{1g}$, $E_g^2$, $A_u^1$ and $A^2{}_{1g}$.*

5. Fig S5 Deposition efficiency on $SiO_2$, Au, $SI_3N_4$, ITO, ebeam patterned and FIB milled substrates

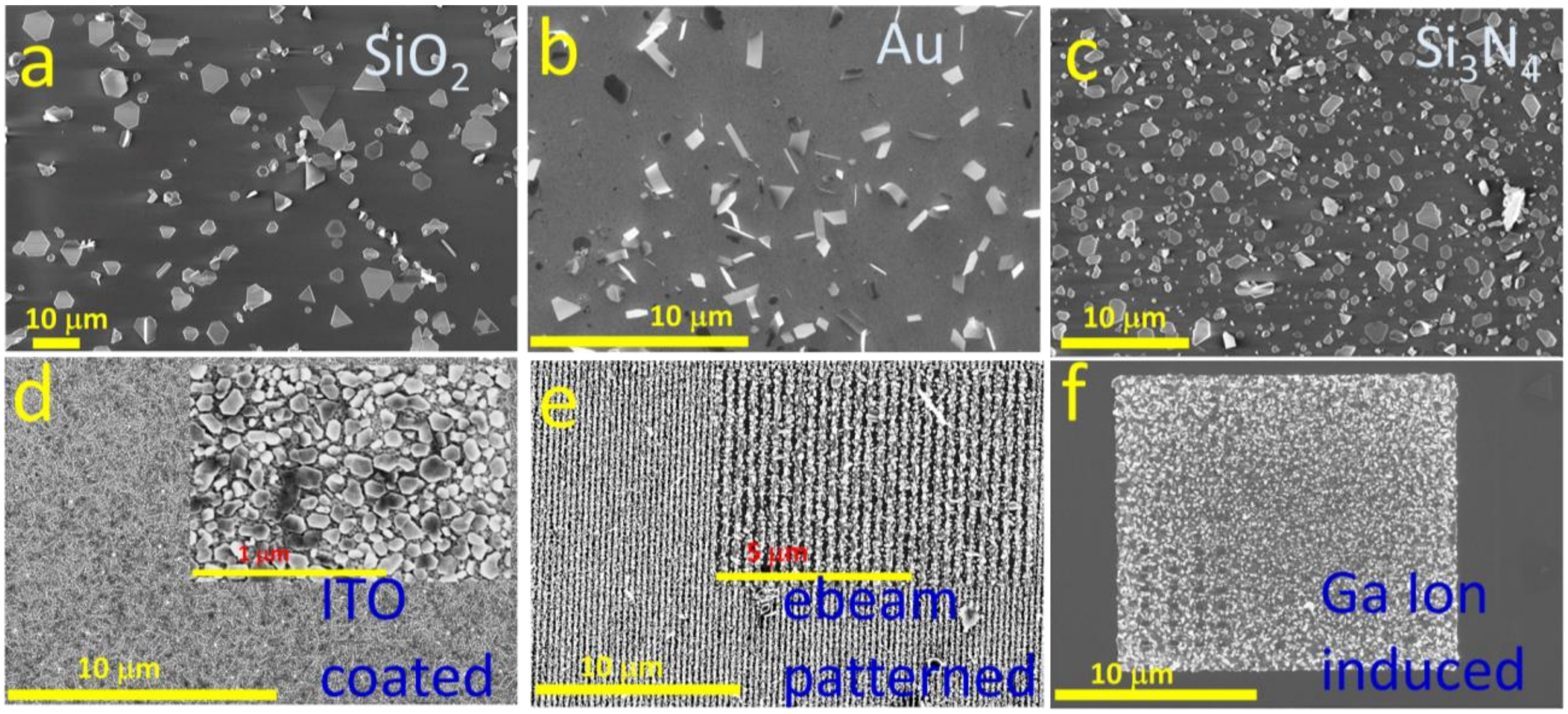

Supplementary information Figure S5 (a-f) represents the different deposition efficiencies on various substrates.

6. Fig S6 Broadspectral photoresponse of different nanostructure devices.

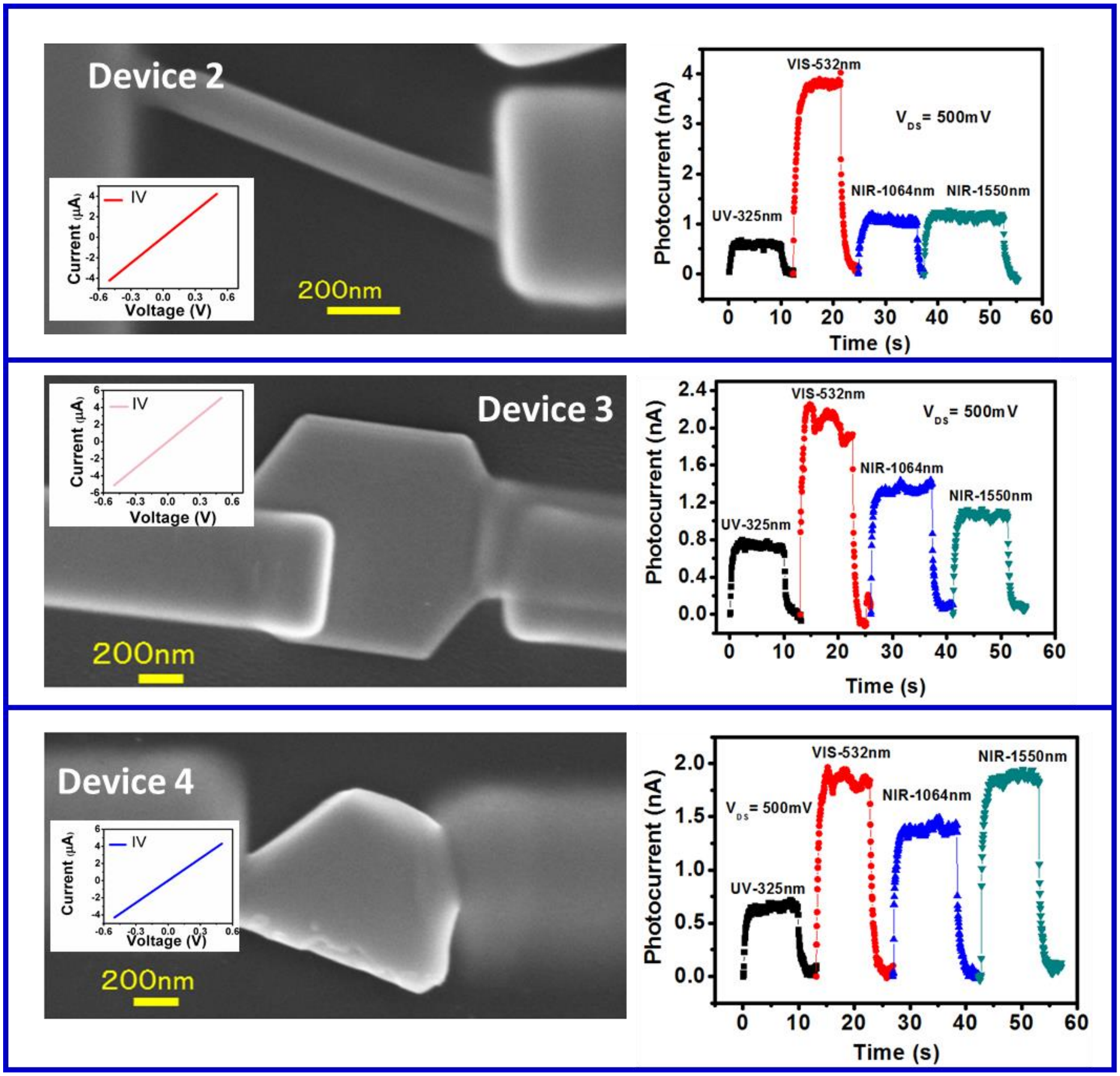

7. Fig S7. Rise and decay time curve fitting

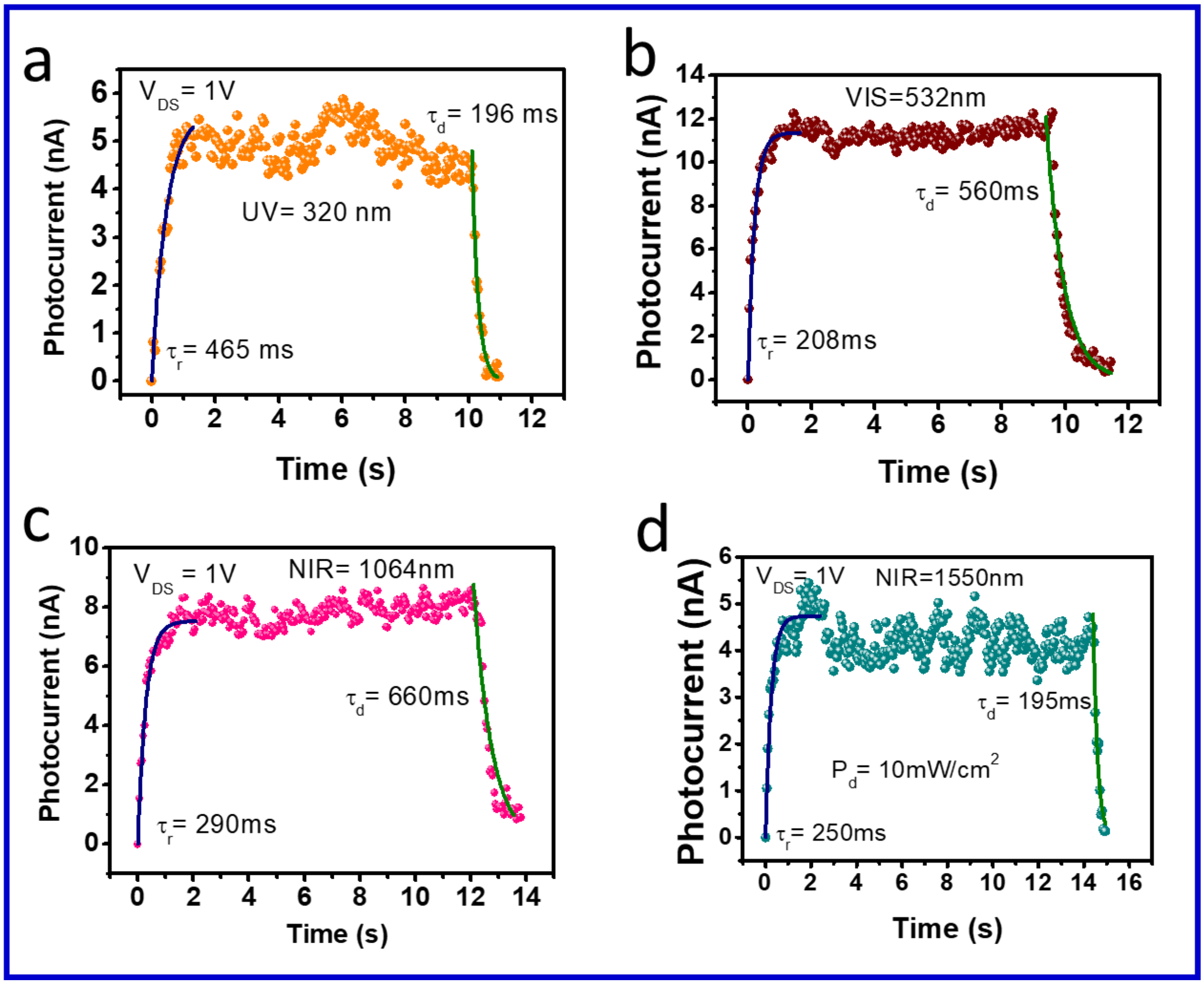

8. Table I. Synthesis comparison of topological insulator nanostructures using various techniques

| Synthesis Method | Advantages | Disadvantages | Ref. |
|---|---|---|---|
| Exfoliation by mechanical (scotch tape)/chemical ( lithium ) technique | Simple, cost effective, clean and pristine samples | Irregular shapes and sizes of the flakes, low deposition yield, no control on the thickness of the exfoliated flakes, poor reproducibility | [1,2] |
| VLS (Vapor liquid solid ) | High quality of nanomaterials and mostly moderate deposition efficiency, synthesis of complex nanostructures | Mostly slow growth, need catalytic particles, unidirectionality, cost-effectiveness, no control over crystal morphologies | [3,4] |
| VS (Vapor solid growth ) | Catalyst free, high quality of nanomaterials,<br>Device friendly substrate, synthesis of complex nanostructures | Deposition efficiency, growth control on specific shapes, cost-effectiveness | [5,6] |
| MOCVD (metal organic chemical vapor deposition) | Concn control of Bi and Te during depostion | Expensive method, chemical based, low deposition efficiency yield | [7,8] |
| MBE (molecular beam epitaxy) | Highest purity in grown samples, film thickness control, easy doping | Highly expensive, high maintenance costs, high vacuum is needed, time consuming, limited for selective substrates and different types of nanostructures. | [9,10] |
| Solvothermal Synthesis | Diverse morphological structures can be synthesized | Chemical and precursor based, not device friendly, uneven solution temp, yield and low purity | [11,12] |
| Pulsed laser deposition (PLD) | Useful as versatile approach, different material combinations | Not useful for large area deposition, uneven coverage, particulate concentration | [13,14] |
| Confined thin film melting deposition :<br>Compared to above techniques our method include i) single step approach (assuming source and deposition substrates are available), ii) growth of nanostructures on device friendly substrates, iii) do not require lattice matching, iv) demonstration of scalable synthesis (on texture substrate ITO), v) simplest way to form hetrostructures , vi) precursor and catalyst free and vii) cost effectiveness.<br>**Future perspective:** The cassette structure based confined melting used in this method just depends on the melting temperature of the material used. There are many novel TMDs (e.g. metal Mo, W and chalcogen atoms S, Se, Te), TI materials beyond the wonder material graphene, those have temperature less than 1200 °C and hence nanostructures of these materials can be synthesized by using this technique in the future where one needs just thin film of such material and hence this method can be used as substitute if compared with the techniques like VLS, CVD or MBE. | | | Our Work |

9. Table II. Photorepsonse properties of topological insulator based nanostructures and thin films.

| **Material** | **λ (nm)** | **R ($AW^{-1}$)** | **D (Jones)** | **Gain/EQE** | **Rise/Decay $\tau_r$ / $\tau_d$, (s)** |
|---|---|---|---|---|---|
| SnTe[15] | 405-3800 | 3.75 | ------- | ----------- | 0.31/085 |
| $Bi_2Se_3$ nanosheets[16] | ------- | 20.48 x $10^{-3}$ | --------- | 8.36 | 0.7/1.48 |
| SnTe/Si[17] | 300-1100 | 2.36 | 1.54 x $10^{14}$ | --- | 2.2 x$10^{-6}$/3.8 x$10^{-6}$ |
| $Sb_2Te_3$ film[18] | 980 | 21.7 | 1.22 x $10^{11}$ | 27.4 | 238.7 / 203.5 |
| $Bi_2Te_3$ NW[19] | 532 | 251 ±0.32 | 4.5 x $10^9$ | 586.15±0.1 | 0.48/0.54 |
| | 325 | 26.82±0.33 | 1.29 x $10^9$ | 102±0.46 | 0.28 / 1.6 |
| Bi film[20] | 370 | 250 x $10^{-3}$ | ---------- | ---------- | 0.9/1.9 |
| $Bi_2Se_3$ film / Si[21] | 808 | 924.2 | 2.38 x $10^{12}$ | 1421 | 0.045/0.047 |
| $Bi_2Te_3$-Graphene[22] | 532 | 35 | ---------- | 83 | 8.7x$10^{-3}$/ 14.8x$10^{-3}$ |
| | 980 | 10 | | 11 | |
| $Bi_2Se_3$ nanowire (NW) [23] | 1064 | 300 | 7.5 x $10^9$ | 350 | 0.550/0.400 |
| $Bi_2Se_3$ (NW)/ Si[24] | 808 | 24.28 | 4.39 x$10^{12}$ | 37.4 | 2.5 x $10^{-6}$ / 5.5 x $10^{-6}$ |
| Polycrystalline $Bi_2Te_3$ / Si [25] | 635 | 1 | 2.5 x $10^{11}$ | ----------- | 0.1/0.1 |
| $WS_2$ -$Bi_2Te_3$[26] | 370-1550 | 30.4 | 2.3 x $10^{11}$ | --------- | 0.020 / 0.020 |